# Dynamic STEM-EELS for single atom and defect measurement during electron beam transformations

Kevin M. Roccapriore,[1] Riccardo Torsi,[2] Joshua Robinson[2], Sergei Kalinin,[3] Maxim Ziatdinov[1,4]

[1] Center for Nanophase Materials Sciences, Oak Ridge National Laboratory, Oak Ridge, TN 37831, USA

[2] Department of Materials Science and Engineering, The Pennsylvania State University, University Park, Pennsylvania 16802, USA

[3] Department of Materials Science and Engineering, University of Tennessee, Knoxville TN, 37916, USA

[4] Computational Sciences and Engineering Division, Oak Ridge National Laboratory, Oak Ridge, TN 37831, USA

ABSTRACT

On- and off-axis electron energy loss spectroscopy (EELS) is a powerful method for probing local electronic structure on single atom level. However, many materials undergo electron-beam induced transformation during the scanning transmission electron microscopy (STEM) and spectroscopy, the problem particularly acute for off-axis EELS signals. Here, we propose and operationalize the rapid object detection and action system (RODAS) for dynamic exploration of the structure-property relationships in STEM-EELS. In this approach, the electron beam is used to induce dynamic transformations creating new defect types at sufficiently small rates and avoiding complete material destruction. The deep convolutional neural networks trained via the ensemble learning iterative training (ELIT) approach are used to identify the defects as they form and perform EELS measurements only at specific defect types. Overall, in this case the EEL spectra are collected only at predefined objects of interest, avoiding measurements on the ideal regions or holes. We note that this approach can be extended to identify new defect classes as they appear, allowing for efficient collection of structure-property relationship data via balanced sampling over defect types.

Layered materials including graphene, dichalcogenides MX$_2$, MXenes, and layered thiophosphates MPS$_3$ have attracted significant attention due to their unique electronic, optical, and mechanical properties.[1–5] Among the multitude of possibilities, important application of layered dichalcogenides is in quantum emitters, which are essential components of quantum technologies, including quantum communication and cryptography.[6] Defects in these materials, such as sulfur vacancies, can lead to the creation of localized electronic states that act as single photon emitters at room temperature.[7,8]

Another promising application of layered dichalcogenides is in catalysis. Transition metal dichalcogenides, such as tungsten diselenide and molybdenum disulfide, have shown potential as electrocatalysts for hydrogen evolution reactions, which are crucial for energy conversion and storage.[9,10] Finally, these materials have shown promise in biological sensing applications. For example, they can be used as biosensors for detecting biomolecules, such as DNA and proteins, due to their high sensitivity to changes in the local environment.[11,12] Defects enable sensing capabilities, as they can create localized binding sites for biomolecules and interactions via the surface states affect the light emitting phenomena or overall conductivity, yielding the detection signal.[13]

Even broader spectrum of functionalities emerges in multilayer systems, where orchestrating the interactions between dissimilar layers via twisting and controlling defect populations opens virtually intractable design space for quantum materials, catalysts, and biological systems.

However, the critical element for understanding these functionalities is the knowledge of the structure-property relationships on a single defect level. After two decades since the broad introduction of aberration corrected Scanning Transmission Electron Microscopy (STEM)[14,15] probing local structure via STEM and extracting local chemical information via core-loss Electron Energy Loss Spectroscopy (EELS) and quasiparticle excitations via low-loss EELS has become almost routine. However, until now these methods could be used only for a limited number of scenarios, where the materials structure remains relatively stable during hyperspectral data acquisition and the defects of interest are sufficiently abundant. This limitation stems from the fundamental operation principle of STEM, based on manual identification of objects of interest by human operator, configuring the EELS experiment over the object of interest, and subsequent manual data analytics. Consequently, the throughput of these experiments is very slow, and the effective dose over the region of interest (ROI) is very high.

The second limitation of this approach is that only stable, already-present defects in the material can be explored. Taken jointly, these factors lead to the fact that very few defects in layered materials are studied in depth via STEM-EELS, and often hours of work of qualified operator led to reliable measurements for only few defect types. Crucially, the measurement of the intended ROI in many samples turns out to represent an entirely different structure than was initially observed due to the electron beam altering the structure and possibly chemistry during the acquisition. This effect is significantly compounded by the recent and perhaps renewed interest in off-axis EELS geometry,[16–18] where acquisition times must be dramatically increased – from milliseconds to seconds – meaning significant electron dose must be imparted to the specimen to collect enough scattered electrons. This collection geometry has been shown to be successful in a relatively few number of cases where the sample is extremely stable under the beam, for instance, vibrational structure of thin film superlattice interfaces,[19] and detecting single-atom vibrational spectroscopy[16,20,21] and more recently even in distinguishing different chemical bonding signals between different silicon dopants in graphene[18], however in the latter, substitutional defects in graphene are widely known to move under the beam.[22–24]

However, we note that the exploration of the structure-property relationships in STEM-EELS can also be performed dynamically. In many layered materials the beam damage effects can lead to slow changes in chemical bonding network, forming novel point and extended defects, nucleating secondary phases, and eventually leading to material degradation. These processes are often slow, and multiple hundreds of STEM images containing defect configurations can be obtained. These systems are unsuitable for the classical grid-based EELS measurements, since for these effective doses are comparable to those during imaging and hence the system is not stationary.

Here, we have developed the rapid object detection and action system (RODAS) workflow combining the STEM imaging at low dose conditions and deep learning to identify and classify defects on the fly, followed by EELS measurements in the designated locations, thus enabling the construction of dynamic library of defect types during the STEM experiment.

## 1. Rapid object detection and action system (RODAS)

Acquisition of EEL spectra at various defect and dopant sites in 2D materials is often a challenge. This is because grid-based spectrum imaging imparts too much electron dose and therefore the desired structure to measure is altered or destroyed. Correspondingly, an intelligent detection and beam control scheme is required to capture the signals. It has been proposed that deep convolutional networks can be used to identify objects of interest in the streaming data. [25,26] However, for classical DCNNs both the object detection and beam control have been elusive and feature detection was too sensitive to hyperparameters for it to be useful on-the-fly. Recently however, ensemble learning allowed to dramatically improve the robustness of model predictions,[27] making real-time analysis possible.[28,29] Within the last few years, controlling the electron beam position (and other microscope parameters) has been made possible within the last few years by manufacturers such as Nion and JEOL by providing Python-based APIs to end users[30,31]. Aside from this, custom control schemes also exist that may be implemented to virtually any instrument where an API is not accessible.

The classical STEM-EELS experiments typically begin with an overview structural image, usually a high-angle annular dark field (HAADF)-STEM image. From this relatively fast scan, an ROI is selected, and a spectrum image acquisition begins, in which a spectrum is collected at a grid of points defined by the ROI, where the pixel size is typically larger than its HAADF counterpart – a result of the increased needed dwell time. In this manner is that the HAADF signal can be synchronously collected, meaning with almost absolute certainty, it is clear what structure has been measured. Aside from decreasing the dwell times as much as possible, a standard method for reducing the total dose and therefore the effect of the beam is to reduce the number of total pixels acquired, i.e., increase the physical size of each pixel. This may be successful if the feature sizes are comparable to the pixel size, however, when probing atomic structures, this option loses its feasibility. One final approach might be to use pixel sizes that are appropriate for realizing structure, but instead of acquiring a single spectrum image with pixel time $t$, instead acquire $n$ spectrum images with pixel time $t/n$. Several groups have used this method successfully but to compensate for specimen spatial drift,[17,32] which might be particularly useful at cryogenic temperatures. Ultimately, this and other approaches still impart far too much electron dose to the specimen that prevents an accurate extraction of the spectral signature of specific defect sites which are generally not extremely robust to the beam.

Alternatively, instead of an entire spectrum image, one may acquire a point spectrum at a chosen location in the overview HAADF-STEM image, by manually positioning the electron beam and recording the spectrum. While potentially dose efficient, this approach suffers multiple downsides: first, it is not necessarily repeatable or reliable to position the electron beam by hand. Scan distortions[33] and non-repeatable human movements can result in significant alignment problems. Second, a simultaneous structural image is not obtained, signifying it is unknown if the specimen has been affected by the beam or drifted from the intended probe location.

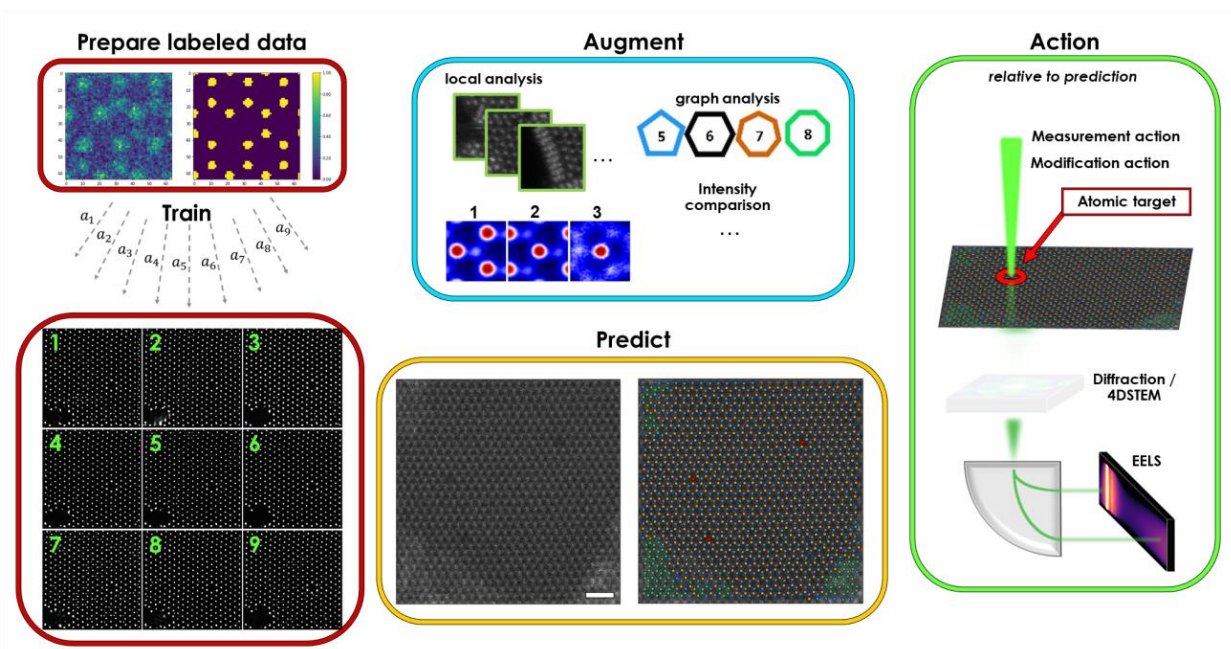

**Figure 1**. Ensemble networks to predict atomic coordinates and classes observed in experiment in real time. Initial set of labeled training data is trained with multiple different random initializations, $a_i$, where ensemble prediction is average of each model prediction within ensemble. To generalize to experimental conditions and multiple classes occurring due to structural changes, ensembles are augmented (retrained) with additional classes encountered during experiment by labeling experimental data (with some manual adjustments), then using it as new training data.

We address these challenges by proposing the workflow that identifies atomic configurations dynamically as they form under the action of electron beam and collects EELS data only on objects of interest. This approach is enabled by a robust feature identification method – a trained ensemble of deep convolutional neural networks (DCNNs) in what is known as ensemble learning iterative training (ELIT)[27]. The ensemble networks allow on-the-fly atomic identification and classification and is the core-enabling aspect of this work and their application here is depicted in **Figure 1**. Note that for full ensemble prediction to be completed in a reasonable timeframe (e.g., within hundreds of milliseconds), one needs to consider appropriate GPU resources for prediction with each model within the ensemble: current off-the-shelf graphics cards can be used for this purpose. Otherwise, a single model from the ensemble can be chosen by the operator, which relaxes the need for expensive GPU computations, at the risk of being slightly less robust. Here, the HAADF-STEM image serves as an input to the ensemble network, which provides atomic coordinates (semantic segmentation) and classes as an output. Combined with flexible probe positioning and access to the EELS camera acquisition, also allow the spectra from detected atomic sites to be acquired in a fully automatic fashion. Note that not all atomic sites will be collected, but generally only the defect sites and usually a small number of regular (non-defect) sites that serve as a basis for comparison.

## 2. Dynamic drift assessment

The primary limitation of the proposed approach stems from the potential probe drift and non-ideality of scanning systems. Even if the HAADF intensity is simultaneously acquired with EELS spectrum, it represents the information from a single pixel only and it is not possible to determine (local) structure from a single pixel. To alleviate this problem, here we acquire a second (fast, and minimal dose) HAADF-STEM image after the spectra are collected, which can then be compared with the original image to determine if a beam-induced change or drift had occurred. Cross-correlation of images is especially difficult or unreliable if there is a beam induced effect causing significant contrast change, therefore in this case we used the motion of all atomic coordinates between frames to detect image shift and change. Alternatively, one may create image patches at each target coordinate in the current and subsequent frame and compare the image patches in time – this assessment may be done by cross correlation. Note that this step can safely be performed post-experiment, implying the comparison method can be modified and finely tuned. At this stage, a pair of images will have been obtained, with a small number (~10) of various spectra collected from specific atomic sites relative to the first image. More data, and therefore statistics, can be collected immediately by simply repeating this process, using the second image to predict atomic coordinates and classes with the ensemble network, acquisition of more EEL spectra at specific atomic sites, followed by a third HAADF-STEM image, and so on. This is repeated as much as possible, where the process (i.e., number of points visited, entire duration, etc.) is dictated by the beam sensitivity of the specimen as well as the sample drift. In other words, the total time for a set of spectrum acquisitions during one step should be sufficiently less than the current drift rate. Similarly, each step should also attempt to use as little as dose as possible, meaning the number of measurement points should be small and simultaneously points should be separated in space to avoid overlapping dose – the latter is usually avoided by using atomic sites as measurement coordinates.

It should be mentioned that the instrumental control via Python commands is limited by the operating system's current demands, and process timing cannot be guaranteed – in other words, the dwell time of the beam is achieved by using a Python command, which can be set equal to the camera exposure time. One might assume that iterating over coordinates in this fashion should perform well, however, due to unpredictable operating system processes, additional delay on the order of hundreds of *ms* can be accumulated with each new positioning command. While the EEL spectrum will correctly represent the appropriate camera dwell time, the beam may ultimately end up dwelling much longer than expected. In the case presented here, dwell times are already >1 second, therefore the extra possible hundred of ms is not a significant change in terms of dose. If one decides to use this approach with much shorter beam dwell times, hardware synchronization must be utilized.

Finally, the sequence of image pairs and measurements are analyzed for drift and changes to the sample after the experiment. It will more than likely be obvious during the experiment when this occurs, but this is generally not a concern, since this approach is fully automated anyway. If in post processing it is discovered that too large a drift or a change to the specimen has occurred during the sequence, it is then truncated at this point and subsequent measurements are discarded, as they are not representative of the intended structure. The overall workflow is represented in **Figure 2**.

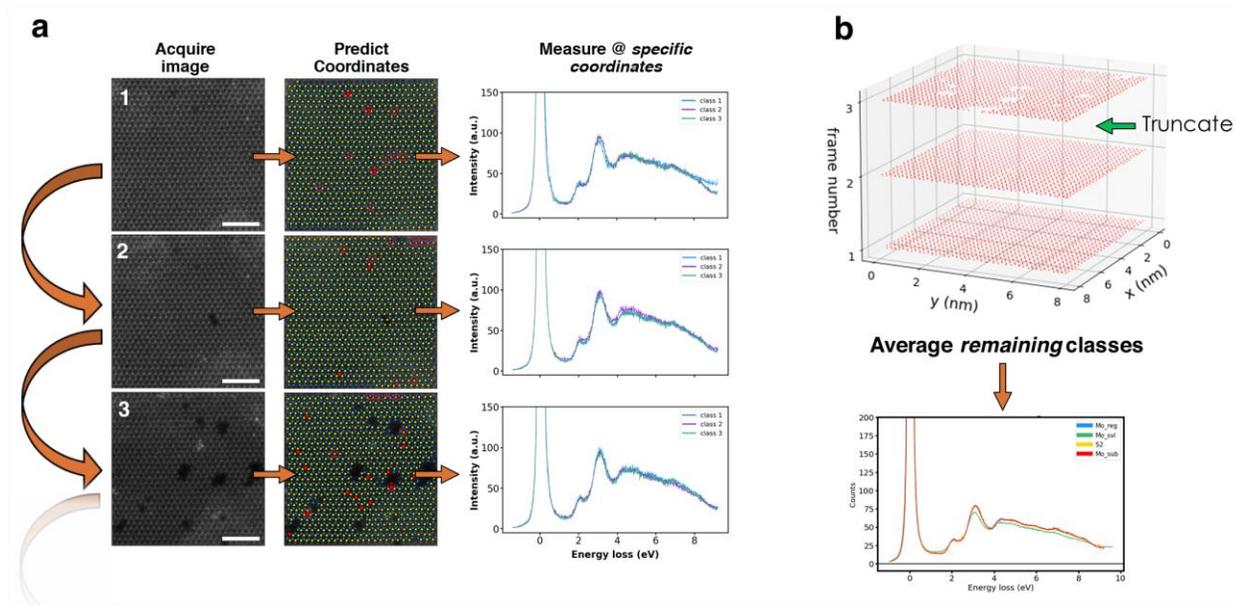

**Figure 2**. Automated site-specific atomic EELS experiment during beam transformations. During a microscope session (a), a HAADF-STEM image is acquired, atomic coordinates are predicted, and several EEL spectra are collected from select different classes and averaged. This is repeated several times to gather more statistics or until the microscopist forces it to end. Post-acquisition analytics (b): atomic coordinates from each frame are used as a metric for both determining if subsequent frames are aligned satisfactorily as well as if the specimen has degraded, transformed, or suffered from contamination. If unsatisfactorily aligned or structural changes occurred, those spectra are discarded, and the remaining classes are averaged.

## 3. Point spectrum EELS and off axis EELS of single atoms and defects

While the proposed framework is universal, here we focus on the applications to STEM-EELS. Several regimes of EELS exist, which exhibit differences in dwell times and electron optic configurations. We consider the following two geometries: on-axis and off-axis EELS. On-axis EELS is the standard geometric configuration where the (scattered) bright field (BF) disc enters the EEL spectrometer aperture. Comparatively, in off-axis EELS, the (scattered) dark field (DF) disc instead enters the spectrometer. The motivation for this is similar logic as to why the dark field signal is used for structural imaging – electrons can undergo primarily either impact scattering or dipole scattering, where the smaller the scattering angle, dipole scattering is strongly favored, and therefore the signal is more delocalized.

In the on-axis EELS configuration (small scattering angles), dipole scattering dominates, which incidentally enables the so-called "aloof" EELS, where the beam can be placed tens or even hundreds of nanometers (depending on energy loss) from a specimen and the electrons still may lose energy via dipole scattering process.[34] While the on-axis EELS signal is several orders of magnitude larger than that of off-axis EELS, the trade-off lies in the (generally) localization of the signals. In general, at core loss energies (>50 eV), the on-axis EEL signal is strongly localized to atoms, so off-axis is not necessarily required for core loss energies. At lower energies losses, and particularly in the regime of optical and vibrational excitations, however, the on-axis signals can become highly delocalized.[35] Consequently, in order to study the relatively low energy vibrational, optical, and electronic excitations of single atoms or defects, it is increasingly necessary (in general) to use an off-axis collection geometry.

The off-axis configuration can be achieved by electrically displacing the aperture with projector lenses, or by using an annular spectrometer aperture. Here we use the former, as it allows simultaneous collection of bright field and dark field information by partially displacing the aperture, such that one portion of the camera receives the bright field information while the remaining portion of the camera receives the dark field information. In this way, the full 2D EELS camera is collected for each measurement and is separated into an on-axis and off-axis portion in post-acquisition analysis, as shown in **Figure 3**. In principle, separation can be done during the experiment, but was not required here.

A point of contention when using the electrically displaced EELS spectrometer aperture is that it results in a known asymmetric annular dark field (aADF) intensity distribution. A portion of the BF disc may now be partially striking the HAADF detector, which is normally restricted to scattering angles on the order of 80-200 mrad. Since a dramatically larger number of electrons is contained in the BF signal relative to higher scattering angle signals, this aADF signal is primarily dominated by the portion of the BF disc that impinges upon the detector. Consequently, the observed structural image more closely resembles annular bright field (ABF) or medium angle annular dark field (MAADF) signals, but this largely depends on the amount of shift to enter this operating geometry. Practically, this has the effect of changing the image contrast. In other words, this type of aADF signal is considered "out of distribution," i.e., the training dataset for most ensemble networks up until this point generally have assumed the image contrast is close to that of what the HAADF signal produces. This simply means that the training data must more closely match experimental conditions, as is the usual case for model training. Finally, to access the low energy excitation regime in STEM-EELS, a form of monochromation is needed. While energy resolutions down to 2.6 meV (at 20 keV primary energy) have been reported,[36] the monochromator itself imparts several additional aberrations to the electron probe. This also can affect image generation, making repeatable atom resolution challenging. In this work, an energy resolution of around 100 meV was used to observe excitations down to about 1 eV as the bandgap in $MoS_2$ is near 1.6 eV – this also allows a larger effective beam current compared to a more strongly monochromated probe.

A Nion monochromated aberration corrected STEM, or MACSTEM, operated at 60 kV equipped with a Dectris ELA detector was used in these automated EELS experiments. Even with electron-counting, direct electron detector (Dectris ELA), the number of electrons collected from a monochromated dark-field signal is extremely small, meaning pixel dwell times on the order of seconds are required to capture a reasonable amount of signal. This is approaching the fundamental limit of collecting every appropriately scattered electron while trying to balance and minimize any beam-induced effect, hence an intelligent positioning scheme is needed to reduce any unnecessary exposure.

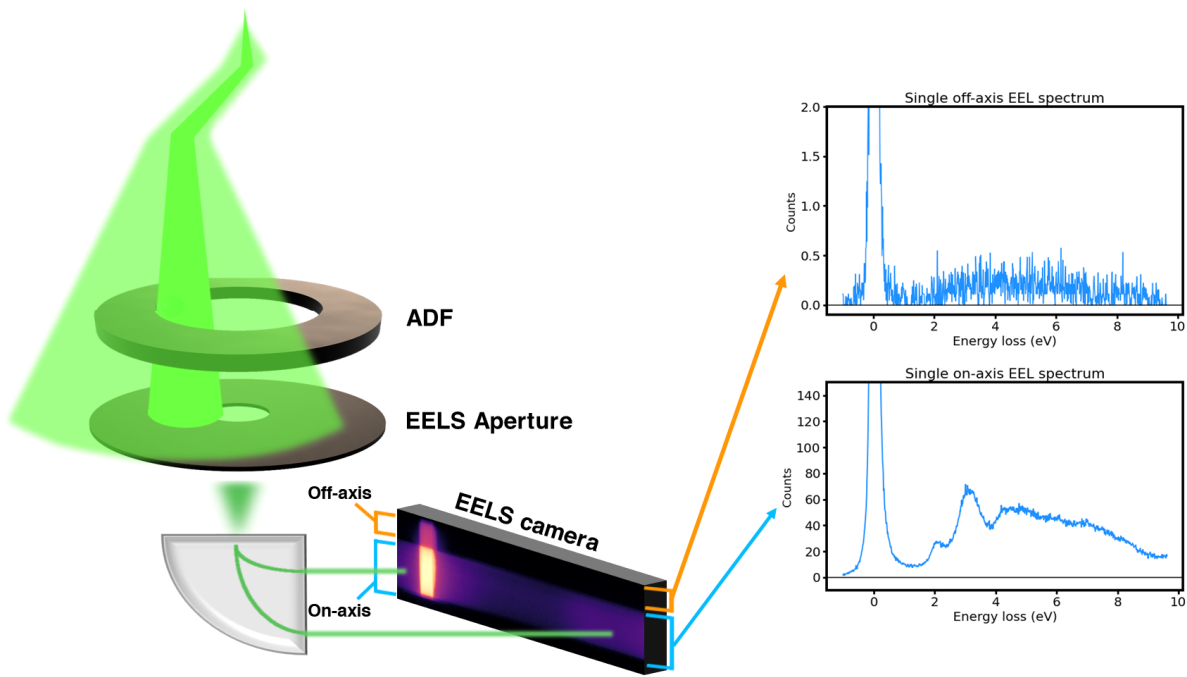

**Figure 3**. On-axis and off-axis: monochromated STEM-EELS. Both on and off axis signals enter the spectrometer and strike the detector, and are easily distinguished in the camera view, therefore two separate vertical regions can be integrated to split the on- and off-axis signals. Background subtraction is not performed, as it is difficult to model the zero loss peak characteristics for a monochromated source, particularly in two dimensions. Averaged raw spectra for each atomic class are shown for both the on- and off-axis EELS scenarios, where little change is observed for on-axis dipole-dominated signals, but the off-axis has much smaller dipole contribution and spectral differences are observed. Note the difference in counts between the two configurations.

As a model system, single-layer 1% V-doped MOCVD-grown $MoS_2$ is explored.[37] It is well known that $MoS_2$ is sensitive to the electron beam in the STEM even at accelerating voltages down to 30kV – this is due to multiple damage mechanisms by the electron beam. Because $MoS_2$ is a semiconductor, both knock-on and radiolysis contribute to the overall damage, however it does remain stable for a handful of HAADF-STEM images to be taken, corresponding to doses approximately on the order of $10^6 - 10^7 \ e^-/Å^2$ before sulfurs are ejected and various defect structures begin to form.[38] Encapsulation of $MoS_2$ with graphene has been shown to protect against damage, however this involves multiple additional steps and introduces more complexity in sample preparation as well as possible different effects on electronic structure.

Upon finding a suitable region of interest, the automated experiment begins by acquiring an overview HAADF, here shown with a field of view (FOV) of 6 or 8 nm, followed by atomic coordinate prediction and classification, and finally, a series of site-specific EEL spectra are collected at several of each class of atom. In this case, we have up to five atomic classes: Mo, mono-S, di-sulfur, V-substitution, or a sulfur vacancy line (SVL) defect, and the algorithm was chosen such that a total acquisition time does not exceed 20 seconds (for drift considerations). Additionally, it seeks to acquire multiple spectra from each class that is detected. Here, we collect the full 2D EELS camera, and segment it into on- and off-axis contributions post-acquisition. As mentioned previously, the spectral data is discarded if either too large of a specified drift has occurred between frames, or structural changes have occurred – in other words, the spectral data is no longer representative of the perceived structures (it dynamically changes through the experiment). After repeatedly performing these automated experiments, the EEL spectra are aligned and averaged separately for each class. These results are presented in **Figure 4**, where the following conclusions can be made. First, there is indeed a difference in the EEL signal arising from the V-substituted Mo site. This can in fact be observed in both on and off axis signals, where the on-axis shows the V spectrum is noticeably larger intensity near the exciton and plasmon energies. On the other hand, the off-axis spectrum shows less intensity – this may be explained by the effective Z number differences between Mo and V, relating to dark field image generation. However, there still appears qualitatively to be slightly different onset slopes, and different features that are not strictly described by difference in Z number. While it was not performed here, vibrational EELS - where energies < 1 eV are considered - is expected to strongly benefit from this workflow.

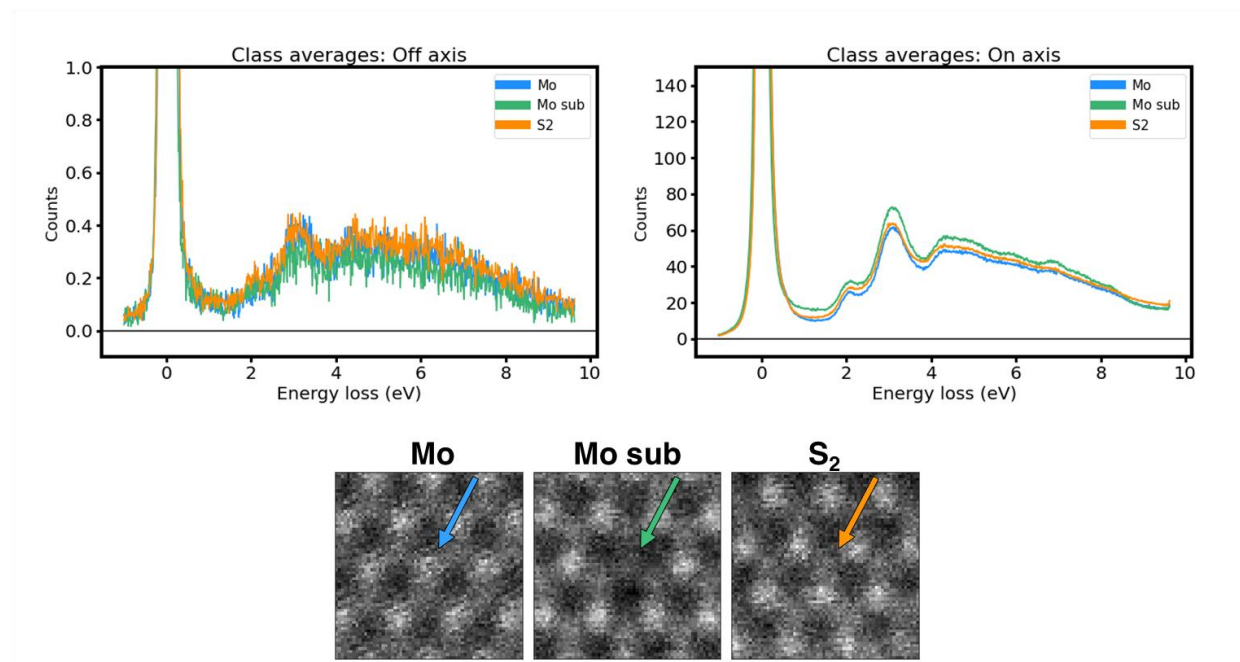

**Figure 4**. Single atom and defect EELS. Collection and averaging of multiple experiments after verification steps Both defect and ordinary atomic classes are collected such that a fair comparison can be made. Both on- and off-axis signals are collected simultaneously and separated by integrating different regions of the detector; off-axis EELS is shown in the left panel, while on-axis EELS is shown in the right panel. Class-averaged image patches for each class are shown on the bottom panel, with arrows indicating from where the spectra were nominally collected.

These experiments can be similarly performed in practically any EELS regime under the realization that the experimental conditions responsible for image generation closely match the image training data for feature detection. Future experiments are required at lower energy (stronger monochromation) to observe any vibrational differences at these scales, and how they might be useful for phononic materials and metamaterials mediated by defect structures. An alternative method could be that instead of acquiring individual point spectra at the detected locations, to acquire a small spectrum image at each target location. Perhaps in this way, a window large enough to capture any structural changes could be used, however additional electrons will be detrimentally deposited into the neighboring structures.

To summarize, here we propose and realize approach for the exploration of the structure property relations complex systems based on the dynamic STEM-EELS measurements of the materials undergoing beam-induced transformations. The classical approaches aim to minimize the beam damage via lowering beam energy or doses. Here, we identify the defects as they form and collect only the data from the defects of interest. Practically, this approach can be trivially extended to prioritize exploration of certain defect classes, focus on emergence of new defects, or balance sampling across possible structural units.

It is again noted that these experiments are not limited to STEM-EELS, and is easily adapted to 4D-STEM (i.e., differential phase contrast, nanobeam electron diffraction, etc.), cathodoluminescence, EDX analysis, etc., and is especially useful in experiments where dwell times are long enough to cause undesirable beam-induced changes. While the primary focus of this work is to enable studying sensitive defects and structures at the *atomic* scale, in principle it is not limited to these sometimes difficult-to-reach scales, or even the STEM, but with the appropriate training and accessibility to system controls, is applicable to a wide variety of microscope modalities.

**Acknowledgements**
This research is sponsored by the INTERSECT Initiative as part of the Laboratory Directed Research and Development Program of Oak Ridge National Laboratory, managed by UT-Battelle, LLC, for the US Department of Energy under contract DE-AC05-00OR22725. The STEM experiments were supported by the U.S. Department of Energy, Office of Science, Basic Energy

Sciences, Materials Sciences and Engineering Division and Oak Ridge National Laboratory's Center for Nanophase Materials Sciences (CNMS), a U.S. Department of Energy, Office of Science User Facility. R.T., and J.A.R. acknowledge funding from NEWLIMITS, a center in nCORE as part of the Semiconductor Research Corporation (SRC) program sponsored by NIST through award number 70NANB17H041 and the Department of Energy (DOE) through award number DESC0010697.